\documentclass[amssymb,amsmath,twocolumn,pra,showpacs,superscriptaddress]{revtex4-1}

\usepackage{epsfig}
\usepackage{pstricks}
\usepackage{color}
\usepackage{bbm}
\usepackage{dsfont}
\usepackage[normalem]{ulem} 
\usepackage{braket}

\usepackage{epstopdf}
\begin{document}
\title{An almost convincing scheme for discriminating  the preparation basis of quantum ensemble and why it will not work}

\author{Sandeep K \surname{Goyal}}
\email{sandeep.goyal@ucalgary.ca}
\affiliation{IQST, Department of Physics and Astronomy, University of Calgary, Calgary, Canada, T2N~1N4.}

\author{Rajeev \surname{Singh}}
\affiliation{The Institute of Mathematical Sciences, CIT Campus, Taramani, Chennai, 600~113.}
\affiliation{Max-Planck-Institut f\"ur Physik komplexer Systeme, N\"othnitzer Stra\ss e 38, Dresden, Germany.}

\author{Sibasish \surname{Ghosh}}
\affiliation{Quantum Science Center, The Institute of Mathematical Sciences, Chennai, 600~113.}

\begin{abstract}
Mixed states of a quantum system, represented by density operators, can be decomposed as a statistical mixture of pure states in a number of ways where each decomposition can be viewed as a different preparation recipe.
However the fact that the density matrix contains full information about the ensemble makes it impossible  to estimate the preparation basis for the quantum system. Here we present a measurement scheme to (seemingly) improve the performance of unsharp measurements. We argue that in some situations this scheme is capable of providing statistics from a single copy of the quantum system, thus making it possible to perform state tomography from a single copy. One of the byproduct of the scheme is a way to distinguish between different preparation methods used to prepare the state of the quantum system. However, our numerical simulations disagree with our intuitive predictions. We show that a counter-intuitive property of a biased classical random walk is responsible for the proposed mechanism not working.
\end{abstract}
\maketitle
\section{Introduction}
We learn in introductory quantum mechanics that the state of a quantum system is described by a vector in a suitable Hilbert space~\cite{Merzbacher,Sakurai}. However such a description is possible only if we have perfect knowledge about the system allowed within quantum theory, in which case the system is said to be in a {\em pure} state. If, on the other hand,  there is some ambiguity about its state, we can no longer attribute a pure state to the system. Such a situation can be described by a statistical mixture of pure states, represented by a density matrix, and the system is  said to be in a {\em mixed} state \cite{,Von-Neumann1927,Fano1957,Schluter1982}.
One can consistently reformulate all the postulates of quantum theory in terms of density matrices and obtain a unified description for both pure and mixed states \cite{Nielsen}.

To understand the context of the present work, we need to familiarize ourselves with a particular aspect of density matrices which is qualitatively different from state vectors---namely, physically different statistical mixtures can give rise to the same density matrix. For example, the equal mixture of spin-up and spin-down particles in the $z$ direction as well as spin-up and spin-down in the $x$ direction results in the same density matrix given by $\mathds{1}/2$.  In fact there are infinitely many ways to decompose a density matrix corresponding to a mixed state and each of them can be thought of as a preparation method \cite{Nielsen}. 

Implicit within the density matrix description of quantum theory is the idea that different preparation methods can not be distinguished, i.e., no physical process can reveal which method was used to prepare a given statistical mixture. In this paper we present a measurement scheme that would appear to violate this idea; i.e., we appear to distinguish between two preparation methods. This measurement scheme uses unsharp measurements  that are also known as positive operator valued measurements (POVMs) \cite{Aharonov1990,Wiseman, Wiseman2001,Gillett2010}. We present a special class of POVMs  that can cause the recurrence of the original state of the quantum system upon repeated measurements. This phenomenon is known as {\em measurement reversal}~ \cite{Royer1994,Amri2011,Cheong2012,He2013,Li2013,Zhi2014,Xiao2013,Li2013,Xiao2014}, where the partially collapsed state of the quantum system (after an unsharp measurement) is reversed back to the original state by performing additional unsharp measurements on it. Measurement reversal has been seen in a number of experiments, both in optical systems as well as in superconducting qubits  \cite{Katz2008,Kim2009,Kim2011,Korotkov2010,Xu2011}. This property is exploited in the present proposal to engineer an efficient method for state tomography. In some rare situations this method is seemingly capable of yielding an accurate   estimation of the state of a single quantum system. If successful it can be used to estimate the preparation basis of an ensemble. There would be severe consequences if this were really possible but we  show that the situation is rescued by a somewhat counter-intuitive property of classical random walks that prove our intuitive argument wrong. We want to emphasize that the purpose of this article is not to mislead people into believing that single copy quantum state tomography is possible but to stimulate further discussions of the measurement postulate  of quantum mechanics.

The article is organized as follows. In Sec.\,\ref{sec-POVM} we revisit the concept of unsharp measurements. Here we discuss  POVMs  and  state tomography using unsharp measurements. In Sec.\,\ref{prep-method} we introduce the concept of measurement reversal and show that it  can be used to estimate the state and hence the preparation bases, from a single quantum system. Details of numerical simulations and results are given in Sec.\,\ref{Num-exp}. In this section we also present an apparent  paradox which arises from the disagreement between our intuition and the numerical results.   We conclude with a summary of results and discussion in Sec.\,\ref{conc}.

\section{Unsharp measurements (POVM)}\label{sec-POVM}
According to the measurement postulate of quantum mechanics,  upon measurement the state of a quantum system will be projected onto one of the eigenstates of the operator corresponding to the observable being measured. Projection onto eigenstates can be represented by projection operators, which for the measurement of  spin of a spin-half particle in the $\hat{n}$ direction read
\begin{align}
P_\pm &= \frac{1}{2}(\mathds{1}_2 \pm \hat{n}\cdot \vec{\sigma}).
\end{align}
Here  $\vec{\sigma} = (\sigma_x,\, \sigma_y,\,\sigma_z)$  is the vector of Pauli spin operators and  $\hat{n}$  is a unit vector representing 
the measured direction of spin in a three-dimensional physical space.

A more general measurement  approach is to make an ancillary system  interact  with the quantum system under observation and perform measurements on the ancillary system afterward \cite{Nielsen}. 
In this  framework  the observable (to be measured) is represented by a collection of positive operators $\{E_i\}$, so-called effects, with  $0 \le E_i \le \mathds{1}$ for all $i$ and $\sum_i E_i = \mathds{1}$.
The $i$-th measurement outcome occurs with probability  ${\rm tr}(\rho E_i)$ and  the state $\rho$ of the measured system transforms as 
$\rho \to \rho'=\frac{1}{{\rm tr}(\rho E_i)}M_{i}\rho M_{i}^\dagger$ \footnote{Unlike projective measurement, knowing the operators $\{E_i\}$ is, in general, not enough  to determine the state of the system after the measurement; we further need to know the operators ${M}_i$'s constituting the POVM elements $\{E_i = M_i^\dagger M_i\}$ because the operators $\{M_i\}$ and $\{UM_i\}$ result in the same POVM elements $\{E_i\}$ for any unitary matrix $U$.
Thus, there is a unitary freedom in choosing the set of the measurement operators $\{M_i\}$. The measurement however is fully specified by the set of measurement operators $\{M_i\}$ in terms of both probabilities of outcome and the state after the measurement.}. Here the measurement operators $M_i$ are related to the effects by $E_i=M_i^\dagger M_i$.

A minimal and complete set of effects $\{E_i\}$ corresponding to the observable $\widehat{O} = \hat{n}\cdot \vec\sigma$ acting on a two-dimensional Hilbert space can be defined as
\begin{align}
E_{\pm} = \frac{1}{2}\left( \mathds{1} \pm \lambda \widehat{O}\right),\label{POVM}
\end{align}
where $0\le \lambda \le 1$.
The parameter $\lambda$ characterizes the strength of the measurement---with $\lambda = 1$ corresponding to the projective measurement and $\lambda = 0$ to the weakest measurement.
 The expectation value of the observable $\widehat O$ is given by
\begin{align}
\langle\, \widehat O\, \rangle &= \frac{1}{\lambda}(\langle\, E_+\, \rangle - \langle\, E_-\, \rangle). \label{expect}
\end{align}

\noindent {\em State tomography:} For the two-level system under consideration, the most general state  is a point in the Bloch sphere which can be written as
\begin{align}
\rho &= \frac{1}{2}\left( \mathds{1} + \vec{r} \cdot \vec{\sigma}\right),
\end{align}
where  $\vec{r}$ is a real three-dimensional vector.
The three real components of the vector $\vec{r}$ are the expectation values of $\sigma_x$, $\sigma_y$, and $\sigma_z$ and they fully characterize the density operator $\rho$.
Thus the state tomography of a two-level system amounts to obtaining these expectation values.
This can be done using Eq.\,\eqref{expect} for the POVM elements corresponding to the measurement in the $x,\,y$, and $z$ directions of the spin.

If we perform measurements on a finite ensemble of  $N$ particles, the probabilities $\langle\, E_+\, \rangle$ and $\langle\, E_-\, \rangle$ are proportional to the number of clicks $N_+$ of $+$ and $N_-$ of $-$, respectively, i.e., $\langle\, E_\pm\, \rangle = N_\pm/N$. If the state $\rho$ is  $(\mathds{1} + \sigma_z)/2$ and the POVM elements are the ones given in Eq.\,\eqref{POVM}, then the probabilities yield $\langle\, E_\pm\, \rangle = N_\pm/N = (1\pm\lambda)/2$, thus, $N_\pm \approx (1\pm\lambda)N/2$. 
However, due to statistical errors we may not always get the exact number of clicks. That will amount to an error in the tomography of the state which will approach zero as we increase the number $N$ to infinity. For a given resolution (error tolerance) it is possible to calculate an appropriate number $N$ which will depend on the measurement strength $\lambda$. 

In the next section we present a scheme for state tomography in which (in rare situations) it is possible to get statistics from a single quantum particle to estimate the state of the system. Although the success probability of such a scenario is very small, together with the traditional methods this scheme provides an improvement in state tomography.

\section{Discriminating Preparation Bases using POVMs}\label{prep-method}

At the heart of improving state tomography are special measurement operators which can cause self-measurement reversal, i.e., the recurrence of the original state of the system upon repeated measurements.  An example of such measurement operators is
  \begin{subequations}
\begin{align}
  M_+ &= \frac{1}{\sqrt{2}} \left(\begin{array}{cc}
\sqrt{1+\lambda} & 0 \\
0 & \sqrt{1-\lambda}
\end{array}\right),\quad  E_+ = M_+^\dagger M_+,\label{eqn-11}\\
M_- &= \frac{1}{\sqrt{2}} \left(\begin{array}{cc}
\sqrt{1-\lambda} & 0 \\
0 & \sqrt{1+\lambda}
\end{array}\right),\quad  E_- = M_-^\dagger M_-.\label{eqn-12}    
\end{align}
\end{subequations}
These measurement operators can be easily constructed by making the quantum system under consideration interact with an ancillary two-level quantum system for a certain time and then performing the projective measurement on the ancillary quantum system. Let the evolution of the joint state $\ket{\psi}\otimes \ket{0}$ of the system plus ancilla be governed by the unitary operator
\begin{align}
  U &= \frac{1}{\sqrt{2}}
        \begin{pmatrix}
          \sqrt{1+\lambda} & -\sqrt{1-\lambda}&0&0\\
          \sqrt{1-\lambda} & \sqrt{1+\lambda} & 0&0\\
          0&0&\sqrt{1-\lambda} & -\sqrt{1+\lambda}\\
          0&0& \sqrt{1+\lambda} & \sqrt{1-\lambda}
        \end{pmatrix}.
\end{align}
We can see that the matrix $U$ acting on $\Ket{\psi}\otimes \Ket{0}$ results in
\begin{align}
  U \Ket{\psi}\Ket{0} = & \frac{1}{\sqrt{2}} \left(\sqrt{1+\lambda} \alpha\Ket{0} + \sqrt{1-\lambda}\beta\Ket{1}\right)\Ket{0}\nonumber \\
  &+ \frac{1}{\sqrt{2}} \left(\sqrt{1-\lambda} \alpha\Ket{0} + \sqrt{1+\lambda}\beta\Ket{1}\right)\Ket{1}.
\end{align}
Now if we perform the measurement in the $\{\Ket{0},\,\Ket{1}\}$ basis on the ancilla, the state of the system collapses to unnormalized states: 
\begin{align}
  \Ket{\psi}_+ &= \frac{1}{\sqrt{2}} \left( \sqrt{1+\lambda} \alpha\Ket{0} + \sqrt{1-\lambda}\beta\Ket{1} \right) \equiv  M_+ \Ket{\psi},\\
  \Ket{\psi}_- &= \frac{1}{\sqrt{2}} \left( \sqrt{1-\lambda} \alpha\Ket{0} + \sqrt{1+\lambda}\beta\Ket{1} \right) \equiv  M_- \Ket{\psi}.
\end{align}

Upon measurement, an arbitrary state $\ket{\psi}$ of the quantum system  transforms as  \footnote{Note that we work with unnormalized states with the square of the norm equal to the probability to obtain the corresponding measurement result.}
\begin{align}
\ket{\psi}\to \left\{\begin{array}{c}
M_+\ket{\psi} \mbox{~with~} p_+ = \bra{\psi}E_+\ket{\psi},\\
M_-\ket{\psi}\mbox{~with~} p_- = \bra{\psi}E_-\ket{\psi}.\end{array}\right.
\label{simulationEqn}
\end{align}
Performing the measurement again on the same system  results in
\begin{align}
\ket{\psi}\to \left\{\begin{array}{c}
M_+\ket{\psi} \to \left\{ \begin{array}{c} M_+^2\ket{\psi},\\M_-M_+\ket{\psi},\end{array}\right.\\
M_-\ket{\psi} \to \left\{ \begin{array}{c} M_+M_-\ket{\psi},\\M_-^2\ket{\psi}.\end{array}\right.\end{array}\right.
\end{align}
It is easy to see that $M_+M_- = M_-M_+ = \sqrt{(1-\lambda^2)/4}\mathds{1}$. Therefore, in these cases we will get back the original state $\ket{\psi}$ with probability $(1-\lambda^2)/2$ after two measurements. If we repeat the measurement on the same system  $q$ number of times we obtain
\begin{align}
\ket{\psi} \to \prod_{i = 1}^q M_{s_i} \ket{\psi},
\end{align}
where $s_i \in \{+,-\}$. If the number of $M_+$'s and $M_-$'s are the same in such sequence of measurements, then we will get back the original state $\ket{\psi}$, the probability of this happening being ${q \choose {q/2}} ((1-\lambda^2)/4)^{q/2}$.

\begin{figure*}
\includegraphics[width=16cm]{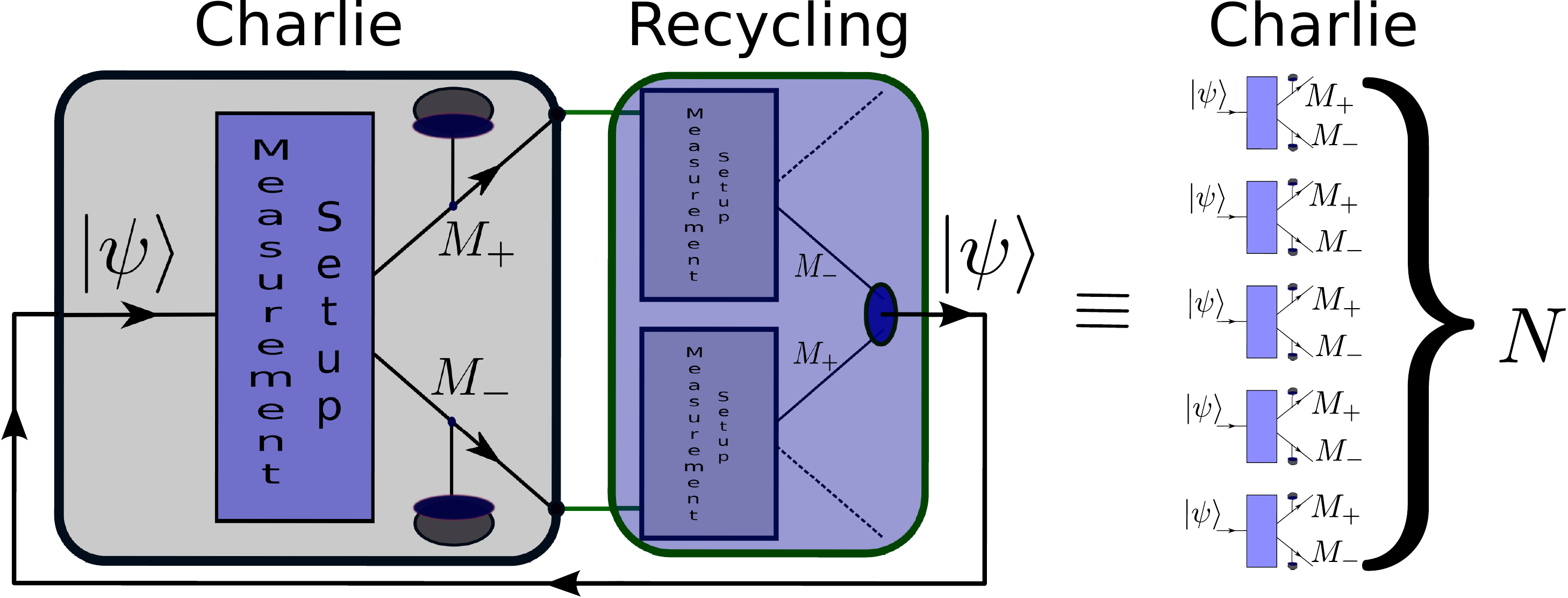}
\caption{ Equivalence between the two situations: (i) where Charlie is performing a POVM measurement on a quantum system in the state $\ket{\psi}$ and afterwards the state is recycled and sent back to Charlie $N$ times, and (ii) where Charlie is provided with $N$ identical quantum systems in the same state. It is apparent from this diagram that the two situations are identical for Charlie.  }
\label{figure}
\end{figure*}

{\em Claim: If we perform a very large number of  measurements on a single quantum system and it regains its original state $N$ times then the whole process is equivalent to the one where we perform measurements on $N$ particles which have identical states. } Consider the following thought experiment: Charlie  is given a single quantum system in state $\ket{\psi}$  to measure the spin in a particular direction. After getting a click in his apparatus, Charlie discards the quantum system and is provided with another quantum system in the same state $\ket{\psi}$. However, to save  resources, the laboratory management decides to recycle the states. To do that they take the discarded system and perform the measurement with the same operators as Charlie multiple times until they get back the original state \footnote{They will know that they have the original state when the number of clicks in $+$  and $-$ are the same (including Charlie's click)}. When their system acquires the original state back they pass it to Charlie to use it in the measurement process. For Charlie, who is unaware of the recycling, there is no difference between a new system and the recycled one, hence fulfilling the claim (see Fig.\,\ref{figure}).   This method of state tomography along with the traditional method  improves the accuracy by effectively increasing the number of copies of the system used for tomography. 

{\em Since the method described above is capable of providing the knowledge about the state of a system from a single copy, one can use it to estimate the preparation basis}. This is done as follows: for simplicity let us assume that the state of the ensemble of the two-level system is $\rho = \mathds{1}/2$. Also let us assume that the preparation basis is the eigenbasis of one of the Pauli operators $\sigma_z$ or  $\sigma_x$. Therefore, $\rho = (\ket{0}\bra{0} + \ket{1}\bra{1})/2$ or $\rho = (\ket{+}\bra{+} + \ket{-}\bra{-})/2$ \footnote{{$\ket{0}$ and $\ket{1}$ ($\ket{+}$ and $\ket{-}$) are the two eigenstates of $\sigma_z$ ($\sigma_x$).}}. In other words, half of the quantum systems in the ensemble are in the pure state $\ket{0}$ ($\ket{+}$) while the other half are in the state $\ket{1}$ ($\ket{-}$) if the preparation basis is the eigenbasis of $\sigma_z$ ($\sigma_x$). Therefore, we know that the state of a given system from the ensemble is one of the pure states from the set $\{\ket{0},\,\ket{1},\,\ket{+},\,\ket{-}\}$. Now we perform a large number of measurements with the measurement operators given in Eqs. \eqref{eqn-11} and \eqref{eqn-12} on each  copy in the ensemble and consider the ones which regained their original state (in the measurement process) for a sufficient number ($N$) of times. For each copy, we collect the number of clicks $N_+$ in $+$ and $N_-$ in $-$ progressively after each time the quantum system regains the original state upon repeated measurement on the copy (i.e. in Charlie's laboratory). Note that if the original state is $\ket{0}$ or $\ket{1}$, the difference in the number of clicks in either outcomes, i.e., $n = N_+ - N_-$ will be around $\pm \lambda N$ and hence the result will be a bimodal histogram [see green curve (open squares) in Fig. \ref{ExtremeBias}].  Whereas the difference tends to vanish when the state is $\ket{+}$ or $\ket{-}$ resulting in a unimodal statistics [see blue curve (open circles) in Fig. \ref{ExtremeBias}]. Thus, by calculating the difference $n$ we can discriminate between the preparation bases.

The idea of preparation basis estimation has severe consequences, such as the possibility of superluminal communication.  To illustrate this with our protocol consider a large number of bipartite maximally entangled states $\ket{\Phi^+} = (\ket{00} + \ket{11})/\sqrt{2}$ shared between Alice and Bob. Alice needs to send a message to Bob. She performs projective measurements either in the $\sigma_z$ basis or in the $\sigma_x$ basis on all the systems. As soon as she performs the measurement each of her systems collapse in one of the eigenstates of the measured observable. Now Bob performs the measurement in the way we discussed above. If we assume that Alice and Bob have an infinite supply of this maximally entangled state and Alice performs her measurements in a fixed basis on all the copies, then in Bob's laboratory there will be particles which will be able to achieve a sufficient number of recurrence of its original state and generate  sufficient statistics for the number $n$ (see inset in Fig.~\ref{NoBias}), the difference between clicks in $+$ and $-$ and hence will be able to tell whether Alice performed $\sigma_x$ measurement or $\sigma_z$ measurement. Since Bob can determine whether Alice measured in the $\sigma_x$ basis or the $\sigma_z$ basis, they have communicated one bit of information over an arbitrary distance instantaneously. This way they can communicate faster than the speed of light.

In the next section we present the numerical simulations for the expected results and the result of the exact simulation of the experimental situation. We show that the numerical results do not agree with our expectations and the scheme discussed above in fact does not work. However, the reason behind the disagreement of the two results reveals a surprising property of biased classical random walks, the proof of  which is presented in the Appendix.

\section{Numerical Experiments}\label{Num-exp}

\begin{figure}
\includegraphics[width=7.5cm]{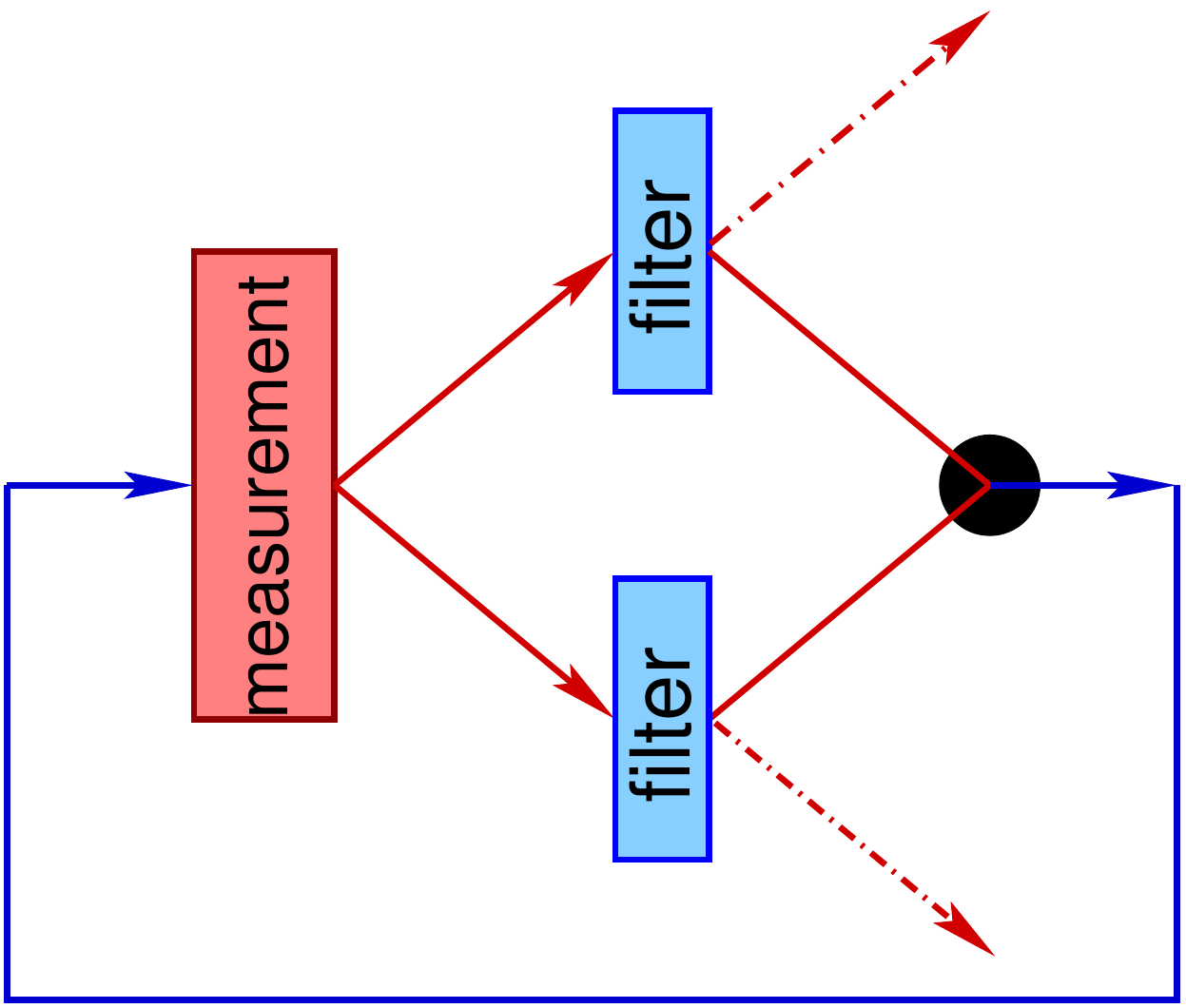}
\caption{ In our Gedanken experiment we run the whole thing in a loop. We start the loop with a measurement in $\{M_+,M_-\}$, followed by a filtering in $\{M_+,M_-\}$ such that the state $M_+\ket{\psi}$ will be mapped to $M_-M_+\ket{\psi}$ and the state $M_-\ket{\psi}$ will be mapped to $M_+M_-\ket{\psi}$. Thus, we feed back  system to the measurement and repeat the process. Now we count how many times $M_+$ clicks in the measurements. As soon as a quantum system completes the loop successfully, it collapses to the state $\ket{\psi}$ back and loses all the information about the history of the path it took to arrive there. Therefore, when we feed the state back to the apparatus, it is exactly like feeding a new system in the state $\ket{\psi}$. Hence the number $N_+$ of the clicks in $M_+$  can give us information even from one single quantum system.  }\label{filter-loop}
\end{figure}

To numerically simulate the protocol discussed above, we start with a
given state $\ket \psi$ and calculate $p_+ = \langle \psi|M_+^{\dagger} M_+|\psi \rangle$. We update $|\psi \rangle$ to $M_+ |\psi \rangle$ with probability $p_+$  and to $M_- |\psi \rangle$ with probability $1-p_+$ and normalize the state afterwards. We repeat the above steps (with updated $p_+$) and keep track of the sequence of $M_+$ and $M_-$ used.  We can map the result of this (numerical) experiment to a one-dimensional classical random walk on a lattice: starting at the origin we move one step to the left when $M_+$ is observed and one step to the right otherwise. In terms of this classical walk the state evolves to the initial state every time the  walker returns to the origin. Therefore, as discussed earlier, recording the result of measurements when the walker is at the origin is equivalent to measurement on a new system from the ensemble. Using the Monte Carlo simulation we can generate very long trajectories using this procedure and count the number of times $M_+$ was observed when the walker was at the origin. To increase the number of returns to the origin we can introduce a filter. This filter discards the trajectories that do not return to the origin at every even step (Fig.~\ref{filter-loop}). In the simulation this is achieved by demanding that the trajectories  return to the origin at every even step; i.e. the odd steps are taken at random with probability $p_+$ above whereas the even steps are the opposite of the previous odd step with probability one. 

\begin{figure}
\includegraphics[width=\columnwidth]{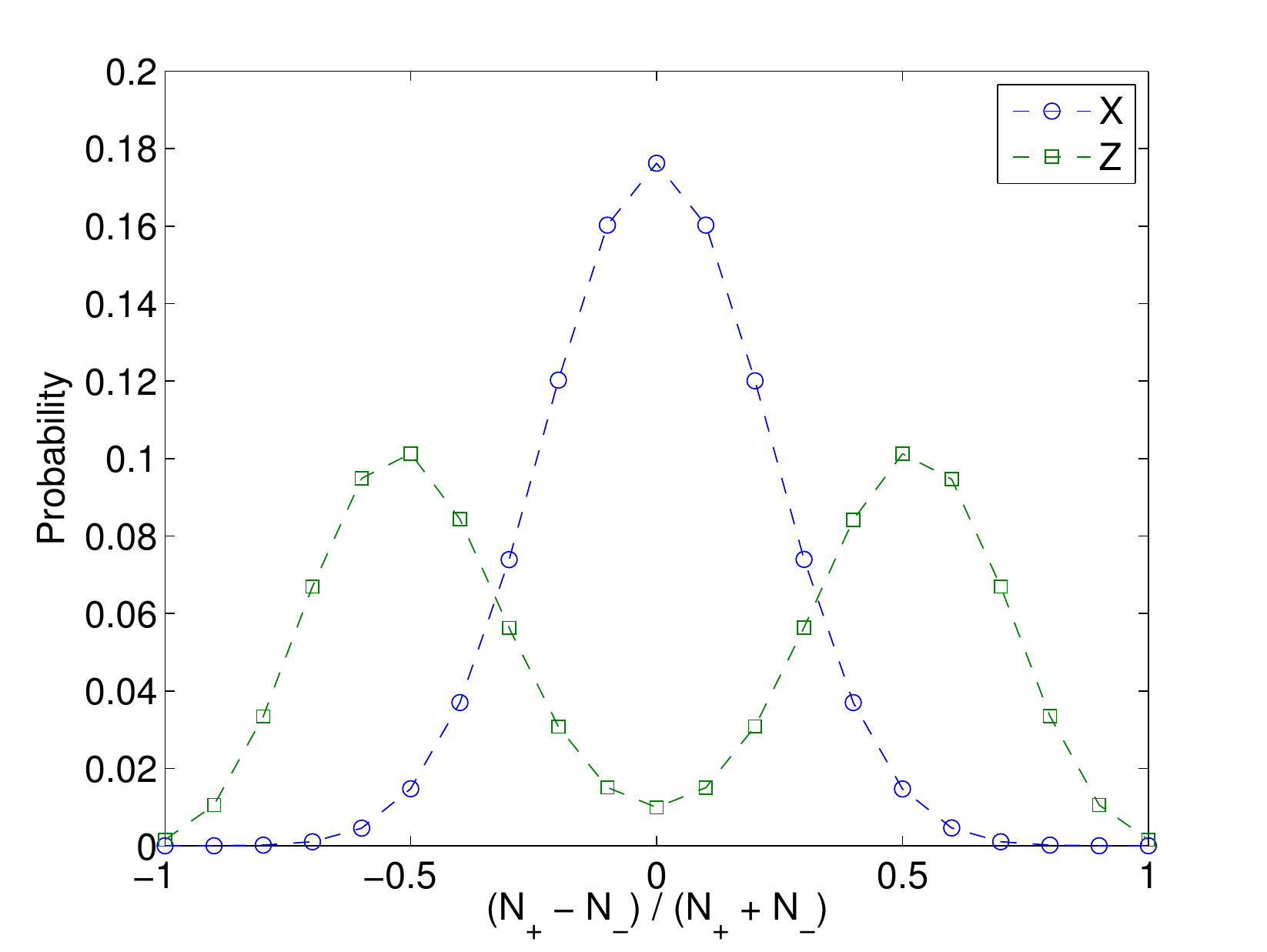}
\caption{Probability of observing the relative number
the two types of output ($M_+$ or $M_-$) in individual trajectories
for the  simulations with the filter as
shown schematically in Fig.~\ref{filter-loop}.
The probabilities for $X$ and $Z$ ensembles are qualitatively
different. }\label{ExtremeBias}
\end{figure}

\begin{figure}
\includegraphics[width=\columnwidth]{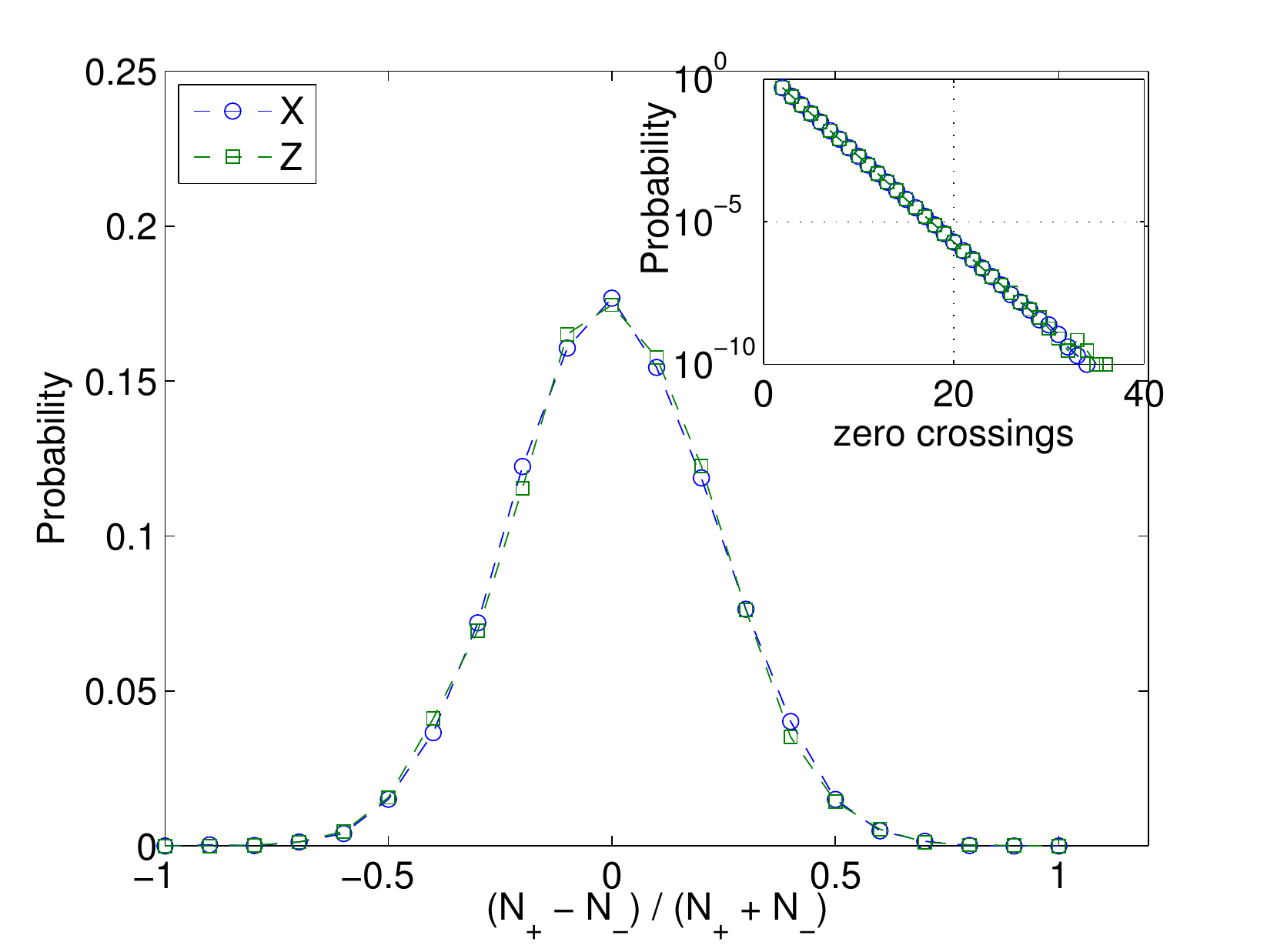}
\caption{Same as Fig.~\ref{ExtremeBias} but for  simulations of the exact experimental protocol (without a filter). Contrary to our expectation both the ensembles show exactly same behavior. The inset shows the probability of the number of zero crossings (in a random
walk sense) for the trajectories, which decreases exponentially with
the number of crossings.}\label{NoBias}
\end{figure}

We are interested in the difference $n = N_+ - N_-$ of the clicks in $M_+$ and in $M_-$ which is the signature for the  bases used in the preparation.
We know that for the basis $\{\ket{0},\,\ket{1}\}$ the number $n$ will be concentrated around either $\lambda N$ or $-\lambda N$. Therefore, $n/N$ should behave as the green curve (open squares) in Fig.\,\ref{ExtremeBias}. Whereas, for the basis $\{\ket{+},\,\ket{-}\}$ the number $n/N$ is concentrated around zero. Thus, it should fall on the blue curve (open circles) in  Fig.~\ref{ExtremeBias}.
We first simulate the system with filters to generate $10^7$ trajectories of length $40$ which cross the origin $20 (= N_+ + N_- = N)$ times.
Thus in our simulation the ensemble contains $10^7$ copies and we make
$40$ measurements on each copy with the constraint that every even
measurement causes reversal.  This constraint results in each copy
acquiring the original state $20$ times. We perform the simulation using
Eq.~\eqref{simulationEqn} and update the state at every measurement.
We show the distribution of the quantity~$n/N = (N_+ -N_-)/N$ in Fig.~\ref{ExtremeBias} which shows a unimodal distribution peaked at $0$ for the $X$ ensemble and a bimodal distribution peaked at $\pm 0.5$ for the $Z$ ensemble (for the choice $\lambda = 0.5$).

To check that we have not introduced any bias by using the filters we also simulate the exact experimental protocol (without filters) to generate $10^{10}$ trajectories of length $1000$.
To compare with Fig.~\ref{ExtremeBias} we use the trajectories which cross the origin exactly $20$ times, which turn out to be only $\sim 10^4$ out of $10^{10}$.
This low rate of generating trajectories with a large number of crossing of the origin make the exact simulation computationally more expensive.
We show the probability of zero crossings in the inset of Fig.~\ref{NoBias}.
By calculating the distribution of the quantity~$n/N$ in Fig.~\ref{NoBias} we find that our protocol does not distinguish the two ensembles.
Thus the numerical simulation of the system with filters, which is in agreement with our intuition, is not accurate in that it does not simulate the experimental situation and introduces bias in the simulation.

The contradiction between the two scenarios can be explained by a rather counter-intuitive property of classical random walks, which is the following.
\newline
{\bf Statement 1:} {\em For a fixed number of zero-crossings even for a biased random walk the trajectories which go to the left and
to the right an equal number of times from the origin are most probable.}
\newline
The proof of this statement can be found in  the Appendix. Note that the numerical experiment corresponding to the simulation of the exact experimental situation (without filters) has a special property for the $Z$ ensemble, namely, the state $|\psi \rangle$ does not change during the evolution. As a result the probability $p_+$ is constant and the mapping to a classical walk is not just a convenience but the $Z$ ensemble trajectories correspond exactly to a biased classical random walk with parameter $p_+$. However, due to  Statement 1 it also results in a unimodal statistics identical to that of the $X$ ensemble.

The analysis presented in this paper suggests an interesting paradox for the measurement on an ensemble of identical pure states (as opposed to a mixture of states discussed so far) which is the following: Charlie is tasked with the measurement of a quantum system in an unknown state $\ket{\psi}$. He is provided with one copy of the state at a time. For him it does not matter how the state is prepared (as long as the state is the same for each copy); he should be able to generate some statistics and estimate the state from a sufficiently large ensemble. However, if the state is prepared by recycling the old state, as was done in Sec. \ref{prep-method}, then Charlie will not be able to get any information about the state otherwise it may cause superluminal communication. Therefore, even though the quantum systems provided to Charlie are identical in every aspect, the way they were prepared will determine the amount of knowledge Charlie can extract from his measurements. 
The paradox becomes more interesting if Charlie knows the state $\ket{\psi}$. In that case by looking at the statistics Charlie can detect whether the quantum systems provided to him are recycled or fresh ones.

\section{Conclusions}\label{conc}

In this article we have presented a measurement scheme to improve the performance of unsharp measurements. We argue that with a very small probability this scheme is capable of providing statistics from a single quantum system, thus making it possible to perform quantum state tomography on a single copy of the quantum system. We show that this result can be used to discriminate different preparation bases used to prepare the state of the quantum system. Therefore, using this scheme one can extract more information about the ensemble than captured by the density operator. One of the severe consequence of this result is the possibility of superluminal communication, thus the result demands more careful study. Numerical simulation of the protocol disagrees with our intuitive understanding. We later provide the resolution for this disagreement. However, an interesting paradox emerges  from this analysis. Since the outcomes of Charlie's experiment depend on the degree of conscious interruptions of the external parties, this paradox seems to point towards the role of a conscious observer in quantum measurements and hence encourage more discussion on the topic.  We hope this paradox will help further the understanding of quantum mechanics, especially the measurement postulate.

\begin{acknowledgments}
S.K.G. would like to thank Thomas Konrad, Rajiah Simon, Christoph Simon, and Ish Dhand for insightful comments. S.K.G. also acknowledges  NSERC for financial support.
\end{acknowledgments}

\appendix

\section{Classical Random Walk}\label{classical-RW}

\begin{figure}
\includegraphics[width=\columnwidth]{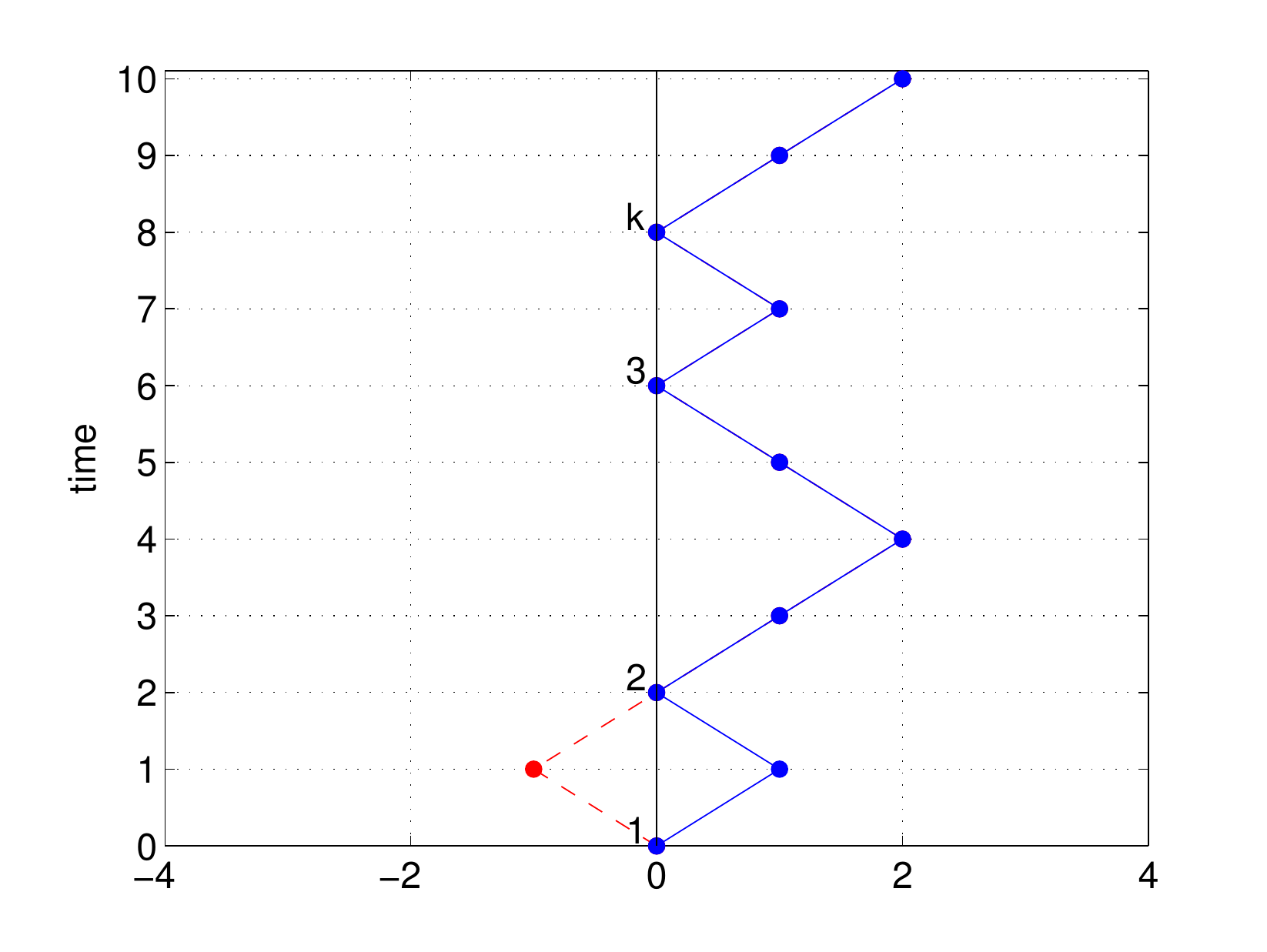}
\caption{ Cartoon showing how to construct possible
walks which go left from the origin once starting from walks that remain
on the right all the time.} \label{ClassicalWalk}
\end{figure}

To prove  Statement 1  let us consider a classical random walk of $T$ steps starting from the origin.
All trajectories that return to the origin at the end of $T$ steps have an equal number of left and right moves and hence have the same probability. Thus the probability of left moves from the origin within this set becomes a purely combinatorial problem.
Let $N_{T,k}(l)$ be the number of trajectories of $T$ steps, returning to the origin $k$ times and going left $l$ times from the origin. We show that $p_{_{T,k}}(l) \equiv N_{T,k}(l) / \sum_i N_{T,k}(i)$ is binomially distributed. In our notation the number of trajectories which never go left from the origin  is $N_{T,k}(0)$.  Using the reflection principle~\cite{Feller}, we find that, for a given trajectory that never goes left from the origin, we get ${k}\choose{1}$ trajectories that go left exactly once by reflecting around the origin one of $k$ parts as can be seen from Fig.~\ref{ClassicalWalk}. Thus the total number of such trajectories is ${k\choose1} N_{T,k}(0)$. Similarly by reflecting $i$ parts of a given trajectory we get ${k}\choose{i}$ trajectories  that go left exactly $i$ times, which amounts to ${k\choose i} N_{T,k}(0)$ such trajectories. Thus we have
\begin{equation}
  N_{T,k}(i) = {k\choose i} N_{T,k}(0)
\end{equation}
which immediately implies that $p_{_{T,k}}(l)$ is a binomial distribution.
Since binomial distribution is symmetric around the center of its range ($k/2$ here), this result implies that indeed our protocol would not be able to distinguish between $X$ and $Z$ ensembles.

Thus we have revealed a very strange property of the classical random walks---
all trajectories of finite length which return to the origin at the end have no information about the bias.
The information about the bias is present in the probability of finding a trajectory that returns to the origin in the end (in our case ${{T}\choose{T/2}} ((1-\lambda^2)/4)^{T/2}$), which is maximum for an unbiased walk and decreases monotonically to zero as the bias is increased.
The trajectories that return to the origin will all have an equal number of left and right moves and hence will be equally likely.
Moreover for every such trajectory there is a dual obtained by reflecting around the origin.
Thus, for example, for every trajectory which moves left every time from the origin there is a trajectory which moves right every time from the origin and has the same probability irrespective of the bias.
This seemingly counter-intuitive property of biased walks can be thought of as the manifestation of the problem of generating fair results from a biased coin~\cite{Neumann1951}.
Since the walker moves left and right from the origin with equal probability in the trajectories which return to the origin at the end, our argument of doing the state tomography from such data would not work.
This particular property of the random walks caused the disagreement between our intuition and the numerical results.


%

\end{document}